\documentstyle[aps,preprint]{revtex}
\begin{document}

\title{  A New Formulation in Lattice Theory}
\author{  {Bo Feng, Jianming Li, Xingchang Song} \\
          {\small { Department of Physics, Peking University, 
          Beijing, 100871, China}}           } 
\maketitle          
 
\begin{center}
\begin{minipage}{120mm}
\begin{center}{\bf Abstract}\end{center} 
{ In this paper we propose a new approach to formulate the field theory 
on a lattice. This approach can eliminate the Fermion doubling problem, 
preserve the chiral symmetry and get
the same dispersion relation for both Fermion and Boson fields. This gives
us the possibility to write down the chiral model ( such as the Weinberg-Salam 
model ) on a lattice.}

\vskip 5mm
\noindent
PACS number(s): 11.15.Ha, 11.30.Rd
\end{minipage}
\end{center}

\par
    It is well-known that the lattice theory suffers from 
the problem of Fermion doubling. To cure from this
Wilson \cite{W1}\cite{W2} added a new term, so-called the ``Wilson term'', 
to the lattice Fermion Lagrangian. This term kills the superfluous components 
of the Fermion field but 
breaks the chiral symmetry and leads to different dispersion relations for 
Fermions and Bosons. Kogut and Susskind\cite{S1}\cite{S2} proposed the 
``stagger model'', but it needs four generations of Fermions at least, 
and deals with the Fermions and Bosons on different footing.
Besides these two popular prescriptions, Drell et al\cite{D} developed another 
approach which preserves the chiral symmetry and at the same time correctly 
counts the number of Fermion states.  Their crucial
point was to introduce a lattice gradient operator which couples all lattice 
sites along a given direction instead of coupling only nearest-neighbor sites.
 
    In this paper, we propose to adopt a new approach in which a new gradient 
operator, in some sence likes the one used by Drell et al, is introduced.
By using our approach, we can eliminate
the Fermion doubling problem and at the same time preserve the chiral symmetry.
Besides these, we can get the same dispersion relation for both Fermions and
Bosons, which will be important if we want to construct the supersymmetry
model on a lattice.
\par
    First let us take a look upon how the Fermion-doubling problem arises. For
free scalar field $\phi(x)$,
the lattice action in the Euclidean space is (for more detail,
see \cite{Cheng} )
\begin{eqnarray}
\label{2}
S(\phi) &=& \sum_n \{ \frac{a^2}{2} \sum_{\mu}(\phi_{n+\mu}-\phi_n)^2
  +\frac{a^4}{2} m^2 \phi_n^2 \} \nonumber \\
 &=& \sum_n \{ \frac{a^2}{2} \sum_{\mu}[(R_{\mu}-1)\phi_n]^2
  +\frac{a^4}{2} m^2 \phi_n^2 \}  ,
\end{eqnarray}
with the definition of the displacement operator $R_{\mu}$ as
\begin{equation}
\label{3}
R_{\mu} \phi_n= \phi_{n+\mu}       .
\end{equation}
Using the formula 
\begin{equation}
\label{4}
\phi_n=\int \frac{d^4k}{(2\pi)^4} e^{ikna} \phi(k),
~~~~-\frac{\pi}{a} \leq k_{\mu} \leq \frac{\pi}{a},~~for~~each~~\mu,  
\end{equation}
we can transfer (\ref{2}) into the momentum space form 
\begin{equation}
\label{5}
S(\phi)=\frac{1}{2} \int \frac{d^4 k}{ (2\pi)^4} \phi(-k) [ \sum_{\mu}
  \frac{4}{a^2} \sin^2(\frac{ak_{\mu}}{2})+m^2] \phi(k)             .
\end{equation}
The dispersion relation read from this equation is
\begin{equation}
\label{6}
S_{\phi}(k)= \sum_{\mu}\frac{4}{a^2} \sin^2(\frac{ak_{\mu}}{2})+m^2       .
\end{equation}

\par
    For Fermion field the lattice action is usually taken as
\begin{equation}
\label{7}
S(\psi)=\sum_n \{ \frac{a^3}{2} \sum_{\mu} \bar{\psi}_n \gamma_{\mu}
   (R_{\mu}-R_{-\mu})\psi_n+ma^4 \bar{\psi}_n \psi_n \} .
\end{equation}
Transfering it into momentum space we get
\begin{equation}
\label{8}
S(\psi)=\int \frac{d^4 k}{ (2\pi)^4} \bar{\psi}(-k) [i\sum_{\mu}
  \gamma_{\mu} \frac{\sin (ak_{\mu})}{a} +m] \psi(k)     .
\end{equation}
From this equation we see the dispersion relation 
\begin{equation}
\label{9}
S_{\psi}(k)= \sum_{\mu}\frac{1}{a^2} \sin^2(ak_{\mu}) + m^2
\end{equation}
is different from (\ref{6}), and a doubling of the Fermionic degrees of freedom 
appears.

\par
    Comparing (\ref{2}) with (\ref{7}) we can see that the difference results 
 from the different definition of the gradient operator $\partial_{\mu}$ 
in bosonic and fermionic cases.

For scalar field we use the replacement 
\begin{equation}
\label{10}
\partial_{\mu} \longrightarrow \frac{R_{\mu}-1}{a} ,
\end{equation}
(hence yielding the D'Alembertian $\Box=\sum_{\mu}\frac{R_{\mu}+R_{-\mu}-2}{a^2}$ 
in the lattice theory effectively).
From some point of view, we can imagine the gradient operator used here  means 
the right-traslation for one unit of lattice spacing $a$ in direction $\mu$. 
So it needs N times for walking all lattice sites in direction $\mu$. 

However for Fermion field to keeping the
Hermitian of the action we use the replacement 
\begin{equation}
\label{11}
\partial_{\mu} \longrightarrow\frac{R_{\mu}-R_{-\mu}}{2a}. 
\end{equation}
We can imagine the action of the gradient operator now means simultaneously
translating the function from a given site to both the right and left ones 
by one unit of lattice spacing $a$ in direction $\mu$.
So only $\frac{N}{2}$ times is needed for walking all points on the lattice. 
This different translation ``speed" explains why there is double-counting in 
formula (\ref{7}) but not in (\ref{5}). It is evident that if one uses the latter
choice of replacement (\ref{11}) instead of the former one (\ref{10}) in (\ref{2}),
there will be double-counting for Boson fields as well. 
So one way of solving the problem 
is to find another replacement of the operator $\partial_{\mu}$ , whose 
action will translate only one unit of lattice spacing at each step
for Fermion field. 

Consulting the definition of the translation operator in the momentun space
\begin{eqnarray}
\label{12}
R_{\mu}\phi_n &=& R_{\mu} [\sum_k e^{ikna} \psi(k)] \nonumber \\
  &=& \phi_{n+\mu}=\sum_k e^{ikna} \psi(k) e^{ik_{\mu}a}  ,
\end{eqnarray}
we can define an operator  $R_{\frac{\mu}{2}}$, corresponding to the 
``half-spacing translation'', as
\begin{eqnarray}
\label{13}
R_{\frac{\mu}{2}} \psi_n &:=& \sum_k e^{ikna} \psi(k) e^{\frac{ik_{\mu}a}{2}};  
\nonumber  \\
R_{-\frac{\mu}{2}} \psi_n &:=& \sum_k e^{ikna} \psi(k) e^{-\frac{ik_{\mu}a}{2}}.
\end{eqnarray}
From the definition (\ref{13}) we obtain
\begin{eqnarray}
\label{b1}
R_{\frac{\mu}{2}}^2 &=& R_{\mu},  \\
\label{b2}
R_{\frac{\mu}{2}} R_{-\frac{\mu}{2}} &=& R_{-\frac{\mu}{2}}R_{\frac{\mu}{2}}
              =I, \\
\label{b3}
\Box &=& \frac{1}{a^2} \sum_{\mu} (R_{\frac{\mu}{2}}-R_{-\frac{\mu}{2}})^2
\end{eqnarray}
Now using the replacement  
\begin{equation}
\label{14}
\partial_{\mu} \longrightarrow
\frac{ R_{\frac{\mu}{2}}-R_{-\frac{\mu}{2}} }{a} 
\end{equation}
in the case of Fermion field
and substituting it into (\ref{7}), we obtain
\begin{equation}
\label{15}
S(\psi)=\sum_n \{ a^3 \sum_{\mu} \bar{\psi}_n \gamma_{\mu}
   (R_{\frac{\mu}{2}}-R_{-\frac{\mu}{2}})\psi_n+ma^4 \bar{\psi}_n \psi_n \}
\end{equation}
and the momentum space form  
\begin{equation}
\label{16}
S(\psi)=\int \frac{d^4 k}{ (2\pi)^4} \bar{\psi}(-k) [2i\sum_{\mu}
  \gamma_{\mu} \frac{ \sin(\frac{ak_{\mu}}{2}) } {a} +m] \psi(k)
\end{equation}
From this we see immediately that there is no double-counting anymore and
the Fermion field has the same dispersion relation as the one for 
scalar field (\ref{6}).
When $m=0$ this new formula (\ref{15}) will preserve the chiral symmetry.

\par
    From the definition (\ref{13}) we can also see that $R_{\frac{\mu}{2}}$ 
and $R_{-\frac{\mu}{2}}$ are not ``local operators". When acting on some 
functions of lattice sites, they not only concern with the nearest neighbor 
sites but also all the sites in direction $\mu$ on the lattice. ( In this 
sence our choice is similar to the one adopted by Drell et al ). However, it 
is evident that the 
combination $\frac{ R_{\frac{\mu}{2}}-R_{-\frac{\mu}{2}} }{a}$ will 
approximate to $\partial_{\mu}$ when the lattice spacing $a \rightarrow 0$. So 
we can call the combination $\frac{R_{\frac{\mu}{2}}-R_{-\frac{\mu}{2}}}{a}$ 
the `` quasi-local operator". The quasi-local feature is the price for getting 
the above mentioned nice characters.
    
    On the other hand, we emphasize that the gradient operator (\ref{14})
we used here is essentially different from that used by Drell et al. Our 
operators $ R_{\frac{\mu}{2}}, R_{-\frac{\mu}{2}}$ and $\partial_{\mu}$ are closely
related to the `` on-site'' operators $ R_{\mu}, R_{- \mu}~ et~ al~ $ through 
Eqs(\ref{b1}),(\ref{b2}) and (\ref{b3}), and share a number of common properties as the `` on-site''
operator (cf (\ref{a1}),(\ref{a2})and (\ref{a3})).
\par
    The next important work is to construct the lattice gauge theory in this
new frame.
    For convenience, let's first  list some useful relations as follows:

\begin{eqnarray}
\label{a1}
R_{\frac{\mu}{2}} (\Phi_n \psi_n) &=& (R_{\frac{\mu}{2}} \Phi_n)
 (R_{\frac{\mu}{2}} \psi_n),   \\
\label{a2}
R_{\frac{\mu}{2}} \psi_n &=& R_{-\frac{\mu}{2}} \psi_{n+\mu}, \\
\label{a3}
[R_{\frac{\mu}{2}} \Phi_n]^+ &=& R_{\frac{\mu}{2}} \Phi_n^+
\end{eqnarray}
and when $a\rightarrow 0$
\begin{eqnarray}
\label{a4}
R_{\mu} \sim  1+a\partial_{\mu},~~~~~~~
R_{\frac{\mu}{2}}  \sim 1+\frac{a}{2} \partial_{\mu}.
\end{eqnarray}

 The gauge invariant action of Fermion field corresponding to (\ref{7}) 
 can be written as
 
\begin{equation}
\label{17}
S(\psi)=\sum_n \{ a^3 \sum_{\mu} \bar{\psi}_n \gamma_{\mu}
   [ U_{n,\frac{\mu}{2}} R_{\frac{\mu}{2}} 
   -U_{n,-\frac{\mu}{2}} R_{-\frac{\mu}{2}})]\psi_n
   +ma^4 \bar{\psi}_n \psi_n \}.
\end{equation}
Here $U_{n,\frac{\mu}{2}}$ is the element of gauge group and can be written
as
\begin{equation}
\label{18}
U_{n,\frac{\mu}{2}}=e^{ i\frac{a}{2}g A_{n,\frac{\mu}{2}}}=exp\{ i\frac{a}{2}g \frac{\lambda^i}{2}
A^i_{n,\frac{\mu}{2}} \}          .
\end{equation}
We want to emphasize that the meaning of two indeces of $U$ are essentially different: 
the first index $n$ denotes the site of lattice point and the second index 
$\frac{\mu}{2}$ denotes the
translation direction and the translation distance. So the operator $R$
acts only on the first index (By this distinction, we can think
$A_{n,\frac{\mu}{2}}$ as the field on lattice site $n$ with an arrow along the  
$\frac{\mu}{2}$ direction, rather than the 
``link variable" in old language) . The local guage transforms of Fermion field
and guage field are given by
\begin{eqnarray}
\label{19}
\psi_n & \longrightarrow &\Phi_n \psi_n ;  \\
\label{20}
U_{n,\frac{\mu}{2}} &\longrightarrow & \Phi_n U_{n,\frac{\mu}{2}}
   [R_{\frac{\mu}{2}} \Phi_n]^+ , \nonumber \\
U_{n,-\frac{\mu}{2}} &\longrightarrow & \Phi_n U_{n,-\frac{\mu}{2}}
   [R_{-\frac{\mu}{2}} \Phi_n]^+ 
\end{eqnarray}
respectively. From (\ref{20}) we can see the guage transform rule of
$U_{n,-\frac{\mu}{2}}$ is same as that of
$[R_{-\frac{\mu}{2}} U_{n,\frac{\mu}{2}}]^+$. So we consider them as the 
same quantity. Then by making use of the Eqs (\ref{a1}) (\ref{a3}) and the Hermitian
of $\lambda^i$ matrices we obtain
\begin{equation}
\label{21}
U_{n,-\frac{\mu}{2}}=exp \{ -i\frac{a}{2}g R_{-\frac{\mu}{2}}A_{n,\frac{\mu}{2}} \} .
\end{equation}

\par
    Now the action of guage field is taken as
\begin{equation}
\label{22}
S(A)=- \frac{8}{g^2} \sum_p tr U_p,~~~~ p\in~~ all~~~plaquettes
\end{equation}
where
\begin{equation}
\label{23}
U_p = U_{n,\frac{\mu}{2}} (R_{-\frac{\mu}{2}} U_{n+\mu,\frac{\nu}{2}})
(R_{-\frac{\mu}{2}} R_{-\frac{\nu}{2}} U_{n+\mu+\nu,-\frac{\mu}{2}} )
(R_{-\frac{\nu}{2}} U_{n+\nu,-\frac{\nu}{2}} )                .
\end{equation}
From (\ref{20}) (\ref{a2}) (\ref{a3}) it is easy to see that $U_p$ is 
gauge invariant. And the above equation can be simplified by using (\ref{21}) 

\begin{eqnarray}
\label{24}
R_{-\frac{\mu}{2}} U_{n+\mu,\frac{\nu}{2}} &=& R_{-\frac{\mu}{2}}
exp \{ i\frac{a}{2}g A_{n+\mu,\frac{\nu}{2}} \}  \nonumber  \\
&=& exp \{ i\frac{a}{2}g R_{-\frac{\mu}{2}} A_{n+\mu,\frac{\nu}{2}} \}
 =exp \{ i\frac{a}{2}g R_{\frac{\mu}{2}}A_{n,\frac{\nu}{2}} \} \nonumber , \\        
R_{-\frac{\mu}{2}} R_{-\frac{\nu}{2}} U_{n+\mu+\nu,-\frac{\mu}{2}} 
&=& R_{-\frac{\mu}{2}} R_{-\frac{\nu}{2}}
     exp \{ -i\frac{a}{2}g R_{-\frac{\mu}{2}}A_{n+\mu+\nu,\frac{\mu}{2}} \}\nonumber  \\ 
&=& exp \{ -i\frac{a}{2}g R_{\frac{\nu}{2}}A_{n,\frac{\mu}{2}} \}   \nonumber , \\
R_{-\frac{\nu}{2}} U_{n+\nu,-\frac{\nu}{2}} 
&=& R_{-\frac{\nu}{2}} 
     exp \{ -i\frac{a}{2}g R_{-\frac{\nu}{2}}A_{n+\nu,\frac{\nu}{2}} \}  \nonumber  \\
&=& exp \{ -i\frac{a}{2}g A_{n,\frac{\nu}{2}} \}   ,
\end{eqnarray}
and then
\begin{equation}
\label{25}
U_p= exp \{ i\frac{a}{2}g A_{n,\frac{\mu}{2}} \}~
     exp \{ i\frac{a}{2}g R_{\frac{\mu}{2}}A_{n,\frac{\nu}{2}} \}
     exp \{ -i\frac{a}{2}g R_{\frac{\nu}{2}}A_{n,\frac{\mu}{2}} \}
     exp \{ -i\frac{a}{2}g A_{n,\frac{\nu}{2}} \}.
\end{equation}
From this expression it can be seen the sum of all ``plaquettes" is equal
to the sum of all different pairs $(n,\mu,\nu)$.

    Now we need to prove when $a \longrightarrow 0$, all the formulae 
we obtained above will
come back to their corresponding continuum form.
First we discuss the action $S(A)$.
Substituting (\ref{a4}) into (\ref{25}) we get
\begin{eqnarray}
\label{26}
U_p  &\sim & exp \{ i\frac{a}{2}gA_{n,\frac{\mu}{2}} \}
exp \{ i\frac{a}{2}g(A_{n,\frac{\nu}{2}}+\frac{a}{2} \partial_{\mu}A_{n,\frac{\nu}{2}}) \}
\nonumber  \\
&  &exp \{ -i\frac{a}{2}g(A_{n,\frac{\mu}{2}}+\frac{a}{2}
       \partial_{\nu} A_{n,\frac{\mu}{2}}) \}
exp \{ -i\frac{a}{2}gA_{n,\frac{\nu}{2}} \}   \nonumber \\
&\sim & exp \{ i\frac{a^2 g}{4} (  \partial_{\mu} A_{n,\frac{\nu}{2} }
                                  -\partial_{\nu} A_{n,\frac{\mu}{2}} +
  ig[A_{n,\frac{\mu}{2}},A_{n,\frac{\nu}{2}}] ) \}\nonumber \\
&\sim &  exp \{ i\frac{a^2 g}{4} F_{n,\mu \nu} \} ,
\end{eqnarray}
with the definition
\begin{equation}
\label{27}
F_{n,\mu \nu}=\partial_{\mu} A_{n,\frac{\nu}{2}}-
\partial_{\nu} A_{n,\frac{\mu}{2}} +
ig[A_{n,\frac{\mu}{2}},A_{n,\frac{\nu}{2}}],
\end{equation}

Then we discuss the continuum form of the kinematic term of the Fermion 
Lagrangian
\begin{equation}
\label{28}
 \bar{\psi}_n \gamma_{\mu}
   [ U_{n,\frac{\mu}{2}} R_{\frac{\mu}{2}}
   -U_{n,-\frac{\mu}{2}} R_{-\frac{\mu}{2}}]\psi_n
\end{equation}
Expanding this equation to the first order of $a$, we get
\begin{eqnarray}
\label{29.a}
& &\bar{\psi}_n \gamma_{\mu} [(1+i\frac{a}{2}gA_{n,\frac{\mu}{2}})(\psi_n+\frac{a}{2}
   \partial_{\mu} \psi_n) \nonumber  \\ 
&  &-(1-i\frac{a}{2}g(A_{n,\frac{\mu}{2}}-\frac{a}{2} 
   \partial_{\mu}A_{n,\frac{\mu}{2}}) )(\psi_n-\frac{a}{2}
   \partial_{\mu} \psi_n)]  \nonumber \\
&\approx& a \bar{\psi}_n \gamma_{\mu} (\partial_{\mu}+
          igA_{n,\frac{\mu}{2}}) \psi_n
\end{eqnarray}
\par
It should be pointed out that, in our scheme, scalar and fermion fields can  
be dealt with in an unified manner. Instead of taking the lattice gradient 
operator as in Eq (\ref{10}), we can also use the new operator in (\ref{14}) 
for scalar field. Then the kinematic part of the Boson Lagrangian should be 
written as 
$$[ (R_{\frac{\mu}{2}}- R_{- \frac{\mu}{2}}) \phi_n]^+
  [ (R_{\frac{\mu}{2}}- R_{- \frac{\mu}{2}}) \phi_n]
$$ which, when transfering to the momentum space, leads to the exactly same formula 
as in Eq (\ref{5}). This means that the dispersion relation (\ref{6}) does not change.
Now if we want to write the gauge invariant version of the scalar action (\ref{2}),
the above kinamatic term has to be changed into

\begin{eqnarray}
\label{30}
[(U_{n,\frac{\mu}{2}}R_{\frac{\mu}{2}}-U_{n,-\frac{\mu}{2}}R_{-\frac{\mu}{2}}
 )\phi_n]^+
[(U_{n,\frac{\mu}{2}}R_{\frac{\mu}{2}}-U_{n,-\frac{\mu}{2}}R_{-\frac{\mu}{2}}
 )\phi_n]
\end{eqnarray}
Using (\ref{20}) (\ref{a1}) (\ref{a3}) it is easy to prove the gauge invariance 
of (\ref{30}).
We also write down the form of (\ref{30}) in the continuum limit as
\begin{eqnarray}
\label{31}
& &(U_{n,\frac{\mu}{2}}R_{\frac{\mu}{2}}-U_{n,-\frac{\mu}{2}}R_{-\frac{\mu}{2}}
 )\phi_n    \nonumber \\
 & \sim &
 (1+i\frac{a}{2}gA_{n,\frac{\mu}{2}})(\phi_n+\frac{a}{2}
\partial_{\mu} \phi_n) 
  -(1-i\frac{a}{2}g(A_{n,\frac{\mu}{2}}-\frac{a}{2}
  \partial_{\mu}A_{n,\frac{\mu}{2}}) )(\phi_n-\frac{a}{2}
\partial_{\mu} \phi_n)  \nonumber \\
&\sim & a (\partial_{\mu}+
igA_{n,\frac{\mu}{2}}) \phi_n
\end{eqnarray}

\par
    Now we can use above results to construct the Weinberg-Salam model
    in lattice theory as an illustration.
    This can be done straightforward by adding
 (\ref{27}) (\ref{13}) (\ref{18}) (\ref{19}) and other common terms together.
 In W-S model, there are two kinds of gauge groups: $SU(2)$ and $U(1)$,
which we denotes as :
\begin{eqnarray}
U_{n,\frac{\mu}{2}}&=&e^{i\frac{a}{2}gA_{n,\frac{\mu}{2}}}=e^{i\frac{a}{2}g \frac{\lambda^i}{2}
A^i_{n,\frac{\mu}{2}}},~~~~~U \in SU(2), \\
V_{n,\frac{\mu}{2}}&=&e^{-i\frac{a}{2}g'B_{n,\frac{\mu}{2}}},~~~V\in U(1).
\end{eqnarray}
Therefore the total lattice action is written as:
\begin{equation}
\label{29}
S_{W-S}=S(\psi)+S(A)+S(B)+S(H)+S(Yukawa)
\end{equation}
where

\begin{equation}
\label{29.1}
S(\psi)=\sum_{\alpha} \sum_n \{ a^3 \sum_{\mu} \bar{\psi}_n^{\alpha}
  \gamma_{\mu} [U_{n,\frac{\mu}{2}}V_{n,\frac{\mu}{2}}R_{\frac{\mu}{2}}
  -U_{n,\frac{-\mu}{2}}V_{n,\frac{-\mu}{2}}R_{\frac{-\mu}{2}}]
  \psi_n^{\alpha} \}
\end{equation}
( here $\alpha$ denotes different kinds of Fermions: left-hand-doublets and
right-hand-singlete of leptons and quarks in three generations.)
\begin{equation}
\label{29.2}
S(A)+S(B)=\frac{-8}{g^2} \sum_p tr U_p+\frac{-8}{g'^2} \sum_p tr V_p,
~~~~ p\in~~ all~~~plaquettes,
\end{equation}

\begin{eqnarray}
\label{29.3}
S(H) &=& \sum_n \{ a^2 \sum_{\mu}
[(U_{n,\frac{\mu}{2}}V_{n,\frac{\mu}{2}}R_{\frac{\mu}{2}}-
U_{n,\frac{-\mu}{2}}V_{n,\frac{-\mu}{2}}R_{\frac{-\mu}{2}}
 )\phi_n]^+
[(U_{n,\frac{\mu}{2}}V_{n,\frac{\mu}{2}}R_{\frac{\mu}{2}} \nonumber \\
&-&U_{n,\frac{-\mu}{2}}V_{n,\frac{-\mu}{2}}R_{\frac{-\mu}{2}}
 )\phi_n]
  +a^4 [\mu^2 \phi_n^+\phi_n +\lambda (\phi_n^+ \phi_n)^2 ]\},
\end{eqnarray}
(here $\phi$ is the Higgs doublets.)
and

\begin{eqnarray}
\label{29.4}
S(Yukawa)&=&\sum_n a^4 \{ [ \bar{l}_{n}^i
\phi_n M_{ij,l} e_{n}^j+h.c]  \nonumber \\
&+&[ \bar{q}_{n,L}^i
\tilde{\phi}_n M_{ij,up} u_{n}^j+h.c]  \nonumber \\
&+& [ \bar{q}_{n,L}^i
\phi_n M_{ij,down} d_{n}^j+h.c]  \}
\end{eqnarray}
(here  $i,j$ denote the three generations.) It can be shown that, with the
helps of Eqs(\ref{26}) (\ref{29.a}) and(\ref{31}), the action (\ref{29}) leads
to the ordinary standard Weinberg-Salam model action in the continuum limit.

\par
    In summary, in this paper we have proposed the replacement of
     $\partial_{\mu}
\longrightarrow \frac{R_{\frac{\mu}{2}}-R_{\frac{-\mu}{2}}}{a}$ in
lattice theory. By doing so, we can eliminate the Fermion-doubling problem,
get the same dispersion relation for both Boson and Fermion fields
and preserve the chiral symmetry. We can also construct the gauge theory
in new frame. Because of these good characters, we can discuss chiral
model  which is difficult to touch upon in normal lattice theory.
 Furthermore we can construct supersymmetry lattice model
in this approach. There will be many interesting problems need to be
considered further and those will be our next works.

\par
\bigskip
\begin{center}
{\bf {\large {Acknowledgments}\ }}
\end{center}

This work is supported in part by the National PanDeng (Clime Up) Plan
and in part by the Chinese National Science Foundation.

\end{document}